\documentclass[12pt]{article}
\usepackage[dvips]{graphicx}

\begin {document}
\begin{flushright}
{\small
SLAC--PUB--10778\\
October 2004\\}
\end{flushright}

\begin{center}
{{\bf\LARGE Novel Aspects of QCD in Leptoproduction}\footnote{Work
supported by Department of Energy contract DE--AC02--76SF00515.}}

\bigskip
{\it Stanley J. Brodsky \\
Stanford Linear Accelerator Center \\
Stanford University, Stanford, California 94309 \\
E-mail:  sjbth@slac.stanford.edu}
\medskip
\end{center}

\vfill

\begin{center}
{\bf\large Abstract }
\end{center}

I review several topics in electroproduction which test fundamental
aspects of QCD.  These include the role of final-state interactions
in producing diffractive leptoproduction processes, the shadowing of
nuclear structure  functions, and target-spin asymmetries. The
antishadowing of nuclear structure functions is shown to be
quark-flavor specific, suggesting that some part of the anomalous
NuTeV result for $\sin^2\theta_W$ could be due to the
non-universality of nuclear antishadowing for charged and neutral
currents. I also discuss the physics of the heavy-quark sea, hidden
color in nuclear wavefunctions, and evidence for color transparency
for nuclear processes. The AdS/CFT correspondence connecting
superstring theory to superconformal gauge theory has important
implications for hadron phenomenology in the conformal limit,
including an all-orders demonstration of counting rules for hard
exclusive processes, as well as determining essential aspects of
hadronic light-front wavefunctions.

\vfill

\begin{center}
{\it Presented at the conference \\
Electron-Nucleus Scattering VIII \\Marciana Marina, Isola d'Elba\\
June 21--25, 2004}
\end{center}
\vfill

\newpage
\section{Introduction}
Although it has been more than 35 years since the discovery of
Bjorken scaling~\cite{Bjorken:1968dy} in
electroproduction~\cite{Bloom:1969kc}, there are still many issues
in deep-inelastic lepton scattering that are only now are being
understood from a fundamental basis in quantum chromodynamics.  This
includes the role of final-state interactions in  producing
diffractive leptoproduction processes, the shadowing of nuclear
structure  functions, and target spin asymmetries.  As I will
discuss, the antishadowing of nuclear structure functions is
quark-flavor specific; this implies that part of the anomalous
NuTeV~\cite{Zeller:2001hh} result for $\sin^2\theta_W$ could be due
to the non-universality  of nuclear antishadowing for charged and
neutral currents.  I also discuss the physics of the heavy-quark
sea, hidden color, and the role of conformal symmetry in hard
exclusive processes. The AdS/CFT correspondence connecting
superstring theory to superconformal gauge theory has important
implications for hadron phenomenology in the conformal limit,
including an all-orders demonstration of  counting rules for hard
exclusive processes, as well as determining essential aspects of
hadronic light-front wavefunctions.

\section{Complications from Final-State Interactions}

It is usually assumed---following the parton model---that the
leading-twist structure functions measured in deep  inelastic
lepton-proton scattering are simply the probability distributions
for finding quarks and gluons in the target nucleon.  In fact, gluon
exchange between the fast, outgoing quarks and the target spectators
effects the leading-twist structure functions in a profound way,
leading to  diffractive leptoproduction processes, shadowing of
nuclear structure  functions, and target spin asymmetries.  In
particular, the final-state  interactions from gluon exchange lead
to single-spin asymmetries in  semi-inclusive deep inelastic
lepton-proton scattering which are not  power-law suppressed in the
Bjorken limit.

A new understanding of the role of final-state interactions in deep
inelastic scattering has recently emerged~\cite{Brodsky:2002ue}.
The final-state interactions from gluon exchange between the
outgoing quark and the target spectator system lead to single-spin
asymmetries (the Sivers effect) in semi-inclusive deep inelastic
lepton-proton scattering at leading twist in perturbative QCD; {\em
i.e.}, the rescattering corrections of the struck quark with the
target spectators are not power-law suppressed at large photon
virtuality $Q^2$ at fixed $x_{bj}$~\cite{Brodsky:2002cx}.  The
final-state interaction from gluon exchange occurring immediately
after the interaction of the current also produces a leading-twist
diffractive component to deep inelastic scattering $\ell p \to
\ell^\prime p^\prime X$ corresponding to color-singlet exchange with
the target system; as discussed below, this in turn produces
shadowing and anti-shadowing of the nuclear structure
functions~\cite{Brodsky:2002ue,Brodsky:1989qz}.  In addition, Paul
Hoyer, Gunnar Ingelman, Rikard Enberg and I have shown that the
pomeron structure function derived from diffractive DIS has the same
form as the quark contribution of the gluon structure
function~\cite{Brodsky:2004hi}. This is discussed in more detail in
Paul Hoyer's contribution to these proceedings.

The final-state interactions occur at a light-cone time $\Delta\tau
\simeq 1/\nu$ after the virtual photon interacts with the struck
quark, producing a nontrivial phase.  Thus none of the above
phenomena is contained in the target light-front wave functions
computed in isolation.   In particular, the shadowing of nuclear
structure functions is due to destructive interference effects from
leading-twist diffraction of the virtual photon, physics not
included in the nuclear light-front wave functions.  Thus the
structure functions measured in deep inelastic lepton scattering are
affected by final-state rescattering, modifying their connection to
light-front probability distributions.  Some of these results can be
understood by augmenting the light-front wave functions with a gauge
link, but with a gauge potential created by an external field
created by the virtual photon $q \bar q$ pair
current~\cite{Belitsky:2002sm}.  The gauge link is also process
dependent~\cite{Collins:2002kn}, so the resulting augmented LFWFs
are not universal.

\section{The Origin of Nuclear Shadowing and Antishadowing}

The shadowing and antishadowing of nuclear structure functions in
the Gribov-Glauber picture is due respectively to the destructive
and constructive interference of amplitudes arising from the
multiple-scattering of quarks in the nucleus.   The effective
quark-nucleon scattering amplitude includes Pomeron and Odderon
contributions from multi-gluon exchange as well as Reggeon
quark-exchange contributions~\cite{Brodsky:1989qz}.  The coherence
of these multiscattering nuclear processes leads to shadowing and
antishadowing of the electromagnetic nuclear structure functions in
agreement with measurements.  Recently, Ivan Schmidt,  Jian-Jun
Yang, and I~\cite{Brodsky:2004qa} have shown that this picture leads
to substantially different antishadowing for charged and neutral
current reactions, thus affecting the extraction of the weak-mixing
angle $\sin^2\theta_W$.  We find that part of the anomalous NuTeV
result for $\sin^2\theta_W$ could be due to the non-universality  of
nuclear antishadowing for charged and neutral currents.  Detailed
measurements of the nuclear dependence of individual quark structure
functions are thus needed to establish the distinctive phenomenology
of shadowing and antishadowing and to make the NuTeV results
definitive.

\section{Light-Front Wavefunctions in QCD}

The concept of a wave function of a hadron as a composite of
relativistic quarks and gluons is naturally formulated in terms of
the light-front Fock expansion at fixed light-front time, $\tau=x
\cdot \omega$.  The four-vector $\omega$, with $\omega^2 = 0$,
determines the orientation of the light-front plane; the freedom to
choose $\omega$ provides an explicitly covariant formulation of
light-front quantization~\cite{cdkm}. The light-front wave functions
(LFWFs) $\psi_n(x_i,k_{\perp_i},\lambda_i)$, with $x_i={k_i \cdot
\omega\over P\cdot \omega}$, $\sum^n_{i=1} x_i=1, $
$\sum^n_{i=1}k_{\perp_i}=0_\perp$, are the coefficient functions for
$n$ partons in the Fock expansion, providing a general
frame-independent representation of the hadron state.

Light-front quantization in the doubly-transverse light-cone
gauge~\cite{Tomboulis:jn,Srivastava:2000cf}  has a number of
advantages, including explicit unitarity, a physical Fock expansion,
exact representations of current matrix elements, and the decoupling
properties needed to prove factorization theorems in high momentum
transfer inclusive and exclusive reactions.

Matrix elements of local operators such as spacelike proton form
factors can be computed simply from the overlap integrals of light
front wave functions in analogy to nonrelativistic Schr\"odinger
theory. For example, one can derive exact formulae for the weak
decays of the $B$ meson such as $B \to \ell \bar \nu
\pi$~\cite{Brodsky:1998hn}  and the deeply virtual Compton amplitude
(DVCS) in the handbag
approximation~\cite{Brodsky:2000xy,Diehl:2000xz}.  An interesting
aspect of DVCS is the prediction from QCD of a $J=0$ fixed Regge
pole contribution to the real part of the Compton amplitude which
has constant energy $s^0 F(t)$ dependence at any momentum transfer
$t$ or photon virtuality~\cite{Brodsky:1971zh,Brodsky:1973hm}. It
arises from the quasi-local coupling of two photons to the quark
current arising from the quark $Z$-graph in time-ordered
perturbation theory of the instantaneous quark propagator arising in
light-front quantization.

One can also define~\cite{Raufeisen:2004dg} a light-front partition
function $Z_{LF}$ as an outer product of light-front wavefunctions.
The deeply virtual Compton amplitude and generalized parton
distributions can then be computed as the trace $Tr[Z_{LF} {\cal
O}],$ where $\cal O$ is the appropriate local operator.  This
partition function formalism can be extended to multi-hadronic
systems and systems in statistical equilibrium to provide a
Lorentz-invariant description of relativistic
thermodynamics~\cite{Raufeisen:2004dg}.

Other applications include two-photon exclusive reactions, and
diffractive dissociation into jets.  The universal light-front wave
functions and distribution amplitudes control hard exclusive
processes such as form factors, deeply virtual Compton scattering,
high momentum transfer photoproduction, and two-photon processes.

One of the central issues in the analysis of fundamental hadron
structure is the presence of non-zero orbital angular momentum in
the bound-state wave functions.  The evidence for a ``spin crisis"
in the Ellis-Jaffe sum rule signals a significant orbital
contribution in the proton wave
function~\cite{Jaffe:1989jz,Ji:2002qa}.  The Pauli form factor of
nucleons is computed from the overlap of LFWFs differing by one unit
of orbital angular momentum $\Delta L_z= \pm 1$.  Thus the fact that
the anomalous moment of the proton is non-zero requires nonzero
orbital angular momentum in the proton wavefunction~\cite{BD80}.  In
the light-front method, orbital angular momentum is treated
explicitly; it includes the orbital contributions induced by
relativistic effects, such as the spin-orbit effects normally
associated with the conventional Dirac spinors.

A number of new non-perturbative methods for determining light-front
wave functions have been developed including discretized light-cone
quantization using Pauli-Villars regularization, supersymmetry, and
the transverse lattice.  One can also project the known solutions of
the Bethe-Salpeter equation to equal light-front time, thus
producing hadronic light-front Fock wave functions.  A potentially
important method is to construct the $q\bar q$ Green's function
using light-front Hamiltonian theory, with DLCQ boundary conditions
and Lippmann-Schwinger resummation.  The zeros of the resulting
resolvent projected on states of specific angular momentum $J_z$ can
then generate the meson spectrum and their light-front Fock
wavefunctions.  For a recent review of light-front methods and
references, see Ref.~\cite{Brodsky:2003gk}.

Diffractive multi-jet production in heavy nuclei provides a novel
way to measure the shape of light-front Fock state wave functions
and test color transparency~\cite{Brodsky:1988xz}. For example,
consider the reaction~\cite{Bertsch:1981py,Frankfurt:1999tq} $\pi A
\rightarrow {\rm Jet}_1 + {\rm Jet}_2 + A^\prime$ at high energy
where the nucleus $A^\prime$ is left intact in its ground state. The
transverse momenta of the jets balance so that $ \vec k_{\perp i} +
\vec k_{\perp 2} = \vec q_\perp < {R^{-1}}_A \ . $ The light-cone
longitudinal momentum fractions also need to add to $x_1+x_2 \sim 1$
so that $\Delta p_L < R^{-1}_A$.  The process can then occur
coherently in the nucleus.  Because of color transparency, the
valence wave function of the pion with small impact separation, will
penetrate the nucleus with minimal interactions, diffracting into
jet pairs~\cite{Bertsch:1981py}. The $x_1=x$, $x_2=1-x$ dependence
of the di-jet distributions will thus reflect the shape of the pion
valence light-cone wave function in $x$; similarly, the $\vec
k_{\perp 1}- \vec k_{\perp 2}$ relative transverse momenta of the
jets gives key information on the derivative of the underlying shape
of the valence pion
wavefunction~\cite{Frankfurt:1999tq,Nikolaev:2000sh}. The
diffractive nuclear amplitude extrapolated to $t = 0$ should be
linear in nuclear number $A$ if color transparency is correct. The
integrated diffractive rate should then scale as $A^2/R^2_A \sim
A^{4/3}$ as verified by E791 for 500 GeV incident pions on nuclear
targets~\cite{Aitala:2000hc}.  The measured momentum fraction
distribution of the jets is consistent with the shape of the pion
asymptotic distribution amplitude, $\phi^{\rm asympt}_\pi (x) =
\sqrt 3 f_\pi x(1-x)$~\cite{Aitala:2000hb}. Data from
CLEO~\cite{Gronberg:1998fj} for the $\gamma \gamma^* \rightarrow
\pi^0$ transition form factor also favor a form for the pion
distribution amplitude close to the asymptotic solution to its
perturbative QCD evolution
equation~\cite{Lepage:1979zb,Efremov:1978rn,Lepage:1980fj}.

\section{Heavy Quark Components of the Proton Structure Function}

In the simplest treatment of deep inelastic scattering, nonvalence
quarks are produced via gluon splitting and DGLAP evolution.
However, in a full theory heavy quarks are multiply-connected to the
valence quarks~\cite{Brodsky:1980pb}. For example, the asymmetry of
the strange and anti-strange distributions in the nucleon is due to
their different interactions with the other quark constituents.
The probability for  Fock states of a light hadron such as the
proton to have an extra heavy quark pair decreases as $1/m^2_Q$ in
non-Abelian gauge theory~\cite{Franz:2000ee,Brodsky:1984nx}.  The
relevant matrix element is the cube of the QCD field strength
$G^3_{\mu nu}.$ This is in contrast to abelian gauge theory where
the relevant operator is $F^4_{\mu \nu}$ and the probability of
intrinsic heavy leptons in QED bound state is suppressed as
$1/m^4_\ell.$ The intrinsic Fock state probability is maximized at
minimal off shellness.  The maximum probability occurs at $x_i = {
m^i_\perp / \sum^n_{j = 1} m^j_\perp}$; i.e., when the constituents
have equal rapidity.   Thus the heaviest constituents have the
highest momentum fractions and highest $x$. Intrinsic charm thus
predicts that the charm structure function has support at large
$x_{bj}$  in excess of DGLAP extrapolations~\cite{Brodsky:1980pb};
this is in agreement with the EMC measurements~\cite{Harris:1995jx}.
It predicts leading charm hadron production and fast charmonium
production in agreement with measurements~\cite{Anjos:2001jr}.   The
production cross section for the double charmed $\Xi_{cc}^+$
baryon~\cite{Ocherashvili:2004hi} and the production of double
$J/\psi's$ appears to be consistent with the dissociation and
coalescence of double IC Fock states~\cite{Vogt:1995tf}.  Intrinsic
charm can also explain the $J/\psi \to \rho \pi$
puzzle~\cite{Brodsky:1997fj}. It also affects the extraction of
suppressed CKM matrix elements in $B$ decays~\cite{Brodsky:2001yt}.
It is thus critical for new experiments (HERMES, HERA, COMPASS) to
definitively establish the phenomenology of  the charm structure
function at large $x_{bj}.$

\section{The Role of Conformal Symmetry in QCD Phenomenology}

The classical Lagrangian of QCD for massless quarks is conformally
symmetric.  Since it has no intrinsic mass scale, the classical
theory is invariant under the $SO(4,2)$ translations, boosts, and
rotations of the Poincare  group, plus the dilatations and other
transformations of the conformal group.  Scale invariance and
therefore conformal symmetry is destroyed in the quantum theory by
the renormalization procedure which introduces a renormalization
scale as well as by quark masses.  Conversely,
Parisi~\cite{Parisi:zy} has shown that perturbative QCD becomes a
conformal theory  for $\beta \to 0$ and zero quark mass.  Conformal
symmetry is thus broken in physical QCD; nevertheless, we can still
recover the underlying features of the conformally invariant theory
by evaluating any expression in QCD in the analytic limit of zero
quark mass and zero $\beta$ function:
\begin{equation}
\lim_{m_q \to 0, \beta \to 0} \mathcal{O}_{QCD} = \mathcal{ O}_{\rm
conformal\ QCD} \ .
\end{equation} This conformal correspondence limit is analogous
to Bohr's correspondence principle where one recovers predictions of
classical theory
from quantum theory in the limit of zero Planck constant.  The
contributions to an
expression in QCD from its nonzero $\beta$-function can be systematically
identified~\cite{Brodsky:2000cr,Rathsman:2001xe,Grunberg:2001bz}
order-by-order in
perturbation theory using the Banks-Zaks procedure~\cite{Banks:1981nn}.

There are a number of useful phenomenological consequences of near
conformal behavior of QCD: the conformal approximation with zero
$\beta$ function can be used as template for QCD
analyses~\cite{Brodsky:1985ve,Brodsky:1984xk} such as the form of
the expansion polynomials for distribution
amplitudes~\cite{Braun:2003rp,Braun:1999te}.  The near-conformal
behavior of QCD is the basis for commensurate scale
relations~\cite{Brodsky:1994eh} which relate observables to each
other without renormalization scale or scheme
ambiguities~\cite{Brodsky:2000cr,Rathsman:2001xe}.  By definition,
all contributions from the nonzero $\beta$ function can be
incorporated into the QCD running coupling $\alpha_s(Q)$ where $Q$
represents the set of physical invariants.  Conformal symmetry thus
provides a template for physical QCD expressions. For example,
perturbative expansions in QCD for massless quarks must have the
form
\begin{equation}
 \mathcal{O} = \sum_{n=0} C_n \alpha^n_s(Q^*_n)
 \end{equation}
where the $C_n$ are identical to the expansion coefficients in the
conformal theory, and $Q^*_n$ is the scale chosen to resum all of
the contributions from the nonzero $\beta$ function at that order in
perturbation theory.  Since the conformal theory does not contain
renormalons, the $C_n$ do not have the divergent $n\!$ growth
characteristic of conventional PQCD expansions evaluated at a fixed
scale.

\section{AFS/CFT Correspondence and Hadronic Light-Front Wavefunctions}
\label{sec:2}

As shown by Maldacena~\cite{Maldacena:1997re}, there is a remarkable
correspondence between large $N_C$ supergravity theory in a higher
dimensional  anti-de Sitter space and supersymmetric QCD in
4-dimensional space-time.  String/gauge duality provides a framework
for predicting QCD phenomena based on the conformal properties of
the AdS/CFT correspondence.

The AdS/CFT correspondence is based on the fact that the generators
of conformal and Poincare transformations have representations on
the five-dimensional anti-deSitter space $AdS_5$  as well as
Minkowski spacetime.  For example, Polchinski and
Strassler~\cite{Polchinski:2001tt} have shown that the power-law
fall-off of hard exclusive hadron-hadron scattering amplitudes at
large momentum transfer can be derived without the use of
perturbation theory by using the scaling properties of the hadronic
interpolating fields in the large-$r$ region of  AdS space.  Thus
one can use the Maldacena correspondence to compute the leading
power-law behavior of exclusive processes such as high-energy
fixed-angle scattering of gluonium-gluonium scattering in
supersymmetric QCD. The resulting predictions for hadron physics
effectively
coincide~\cite{Polchinski:2001tt,Brower:2002er,Andreev:2002aw} with
QCD dimensional counting
rules:\cite{Brodsky:1973kr,Matveev:ra,Brodsky:1974vy}
\begin{equation}
\frac{d\sigma}{dt}(H_1 H_2 \to H_3 H_4) =\frac{
F(t/s)}{s^{n-2}}\end{equation}
where $n$ is the sum of the minimal
number of interpolating fields in the initial and final state. (For
a recent review of hard fixed $\theta_{CM}$ angle exclusive
processes in QCD see reference~\cite{Brodsky:2002st}.)  Polchinski
and Strassler~\cite{Polchinski:2001tt} have also derived counting
rules for deep inelastic structure functions at $x \to 1$ in
agreement with perturbative QCD
predictions~\cite{Lepage:1980fj,Brodsky:1994kg} as well as
Bloom-Gilman exclusive-inclusive duality~\cite{Bloom:1971ye}.

The supergravity analysis is based on an extension of classical
gravity theory in higher dimensions and is nonperturbative.  Thus
analyses of exclusive processes~\cite{Lepage:1980fj} which were
based on perturbation theory can be extended by the Maldacena
correspondence to all orders.  An important point is that the hard
scattering amplitudes which are normally or order $\alpha_s^p$ in
PQCD appear as order $\alpha_s^{p/2}$ in the supergravity
predictions.  This can be understood as an all-orders resummation of
the effective potential~\cite{Maldacena:1997re,Rey:1998ik}.

The superstring theory results are derived in the limit of a large
$N_C$~\cite{'tHooft:1973jz}.  For gluon-gluon scattering, the
amplitude scales as ${1}/{N_C^2}$.   For color-singlet bound states
of quarks, the amplitude scales as ${1}/{N_C}$.  This large
$N_C$-counting, in fact, corresponds to the quark interchange
mechanism~\cite{Gunion:1973ex}.  For example, for $K^+ p \to K^+ p$
scattering, the $u$-quark exchange amplitude scales approximately as
$\frac{1}{u}$\ $\frac{1}{t^2},$ which agrees remarkably well with
the measured large $\theta_{CM}$ dependence of the $K^+ p$
differential cross section~\cite{Sivers:1975dg}.  This implies that
the nonsinglet Reggeon trajectory asymptotes to a negative
integer~\cite{Blankenbecler:1973kt}, in this case, $\lim_{-t \to
\infty}\alpha_R(t) \to -1.$

De Teramond and I have extended the Polchinski-Strassler analysis to
hadron-hadron scattering~\cite{Brodsky:2003px}.  We have also shown
how to compute the form and scaling of light-front hadronic
wavefunctions using the AdS/CFT correspondence in quantum field
theories which have an underlying conformal structure, such as
${\mathcal N} = 4$ super-conformal QCD. For example, baryons are
included in the theory by adding an open string sector in $AdS_5
\times S^5$ corresponding to quarks in the fundamental
representation of  the $SU(4)$ symmetry defined on $S^5$ and the
fundamental and higher representations of $SU(N_C).$ The hadron mass
scale is introduced by imposing boundary conditions at the $AdS_5$
coordinate  $r= r_0 = \Lambda_{QCD} R^2.$ The quantum numbers of the
lowest Fock state of each hadron, including its internal orbital
angular momentum and spin-flavor symmetry, are identified by
matching the fall-off of the string wavefunction  $\Psi(x,r)$ at the
asymptotic $3+1$ boundary.  Higher Fock states are identified with
conformally invariant quantum fluctuations of the bulk geometry
about the AdS background.  The eigenvalues of the 10-dimensional
Dirac and Rarita-Schwinger equations have also been used to
determine the  nucleon and $\Delta$ spectrum in conformal QCD.  The
results are in surprising agreement with the empirical
spectra~\cite{deTeramond:2004qd}.

The scaling and conformal properties of the AdS/CFT correspondence
leads to a hard component of light-front wavefunctions of the
form~\cite{Brodsky:2003px}:
\begin{eqnarray}
\psi_{n/h} (x_i, \vec k_{\perp i} , \lambda_i, l_{z i}) &\sim&
\frac{(g_s~N_C)^{\frac{1}{2} (n-1)}}{\sqrt {N_C}} ~\prod_{i =1}^{n
- 1} (k_{i \perp}^\pm)^{\vert l_{z i}\vert} ~ \nonumber \\[1ex]
&\times&\left[\frac{ \Lambda_o}{ {M}^2 - \sum _i\frac{\vec k_{\perp i}^2 +
m_i^2}{x_i} +
\Lambda_o^2}  \right] ^{n +\vert l_z \vert -1}, \label{eq:lfwfR}
\end{eqnarray}
where $g_s$ is the string scale and $\Lambda_o$ represents the basic
QCD mass scale.  The scaling predictions agree with  perturbative
QCD analyses~\cite{Ji:bw,Lepage:1980fj}, but the AdS/CFT analysis is
performed at strong coupling without the use of perturbation theory.
The near-conformal scaling properties of light-front wavefunctions
lead to a number of other predictions for QCD which are normally
discussed in the context of perturbation theory, such as constituent
counting scaling laws for structure functions at $x \to 1$, as well
as the leading power fall-off of form factors and hard exclusive
scattering amplitudes for QCD processes.

John Hiller, Dae Sung Hwang, Volodya Karmanov, and I have recently
studied the analytic structure of light-front wave functions and its
consequences for hadron form factors using the explicitly
Lorentz-invariant formulation of the front
form~\cite{Brodsky:2003pw} where  the normal to the light front is
specified by a general null vector $\omega^\mu.$ The resulting LFWFs
have definite total angular momentum, are eigenstates of a {\it
kinematic} angular momentum operator, and satisfy all Lorentz
symmetries. They are analytic functions of the invariant mass
squared of the constituents $M^2_0= (\sum k^\mu)^2 = \sum
{k^2_{\perp i}+m^2_i\over x_i}$ and the light-cone momentum
fractions $x_i= {k_i\cdot \omega / p \cdot \omega}$ multiplied by
invariants constructed from the spin matrices, polarization vectors,
and $\omega^\mu.$ These properties can be explicitly verified using
known nonperturbative eigensolutions of the Wick--Cutkosky model.
The dependence of LFWFs on $M^2_0$ also agrees with the conformal
form given above.  The analysis implies that hadron form factors are
analytic functions of $Q^2$ in agreement with dispersion theory and
perturbative QCD.

The leading-twist PQCD predictions~\cite{Lepage:1980fj} for hard
exclusive amplitudes are written in the factorized form as a
convolution of hadron distribution amplitudes $\phi_I(x_i,Q)$ for
each hadron $I$ times the hard scattering amplitude $T_H$ obtained
by replacing each hadron with collinear on-shell quarks with
light-front momentum fractions $x_i = k^+_i/P^+.$ The hadron
distribution amplitudes are obtained by integrating the $n-$parton
valence light-front wavefunctions: $$\phi(x_i,Q) = \int^Q
\Pi^{n-1}_{i=1}  d^2 k_{\perp i} ~ \psi_{\rm val}(x_i,k_\perp).$$
Thus the distribution amplitudes are $L_z=0$ projections of the LF
wavefunction, and the sum of the spin projections of the valence
quarks must equal the $J_z$ of the parent hadron. Higher orbital
angular momentum components lead to power-law suppressed exclusive
amplitudes~\cite{Lepage:1980fj,Ji:2003fw}. Since quark masses can be
neglected at leading twist in $T_H$, one has quark helicity
conservation, and thus, finally, hadron-helicity conservation: the
sum of initial hadron helicities equals the sum of final helicities.
In particular, since the hadron-helicity violating Pauli form factor
is computed from states with $\Delta L_z = \pm 1,$  PQCD predicts
$F_2(Q^2)/F_1(Q^2) \sim 1/Q^2 $ [modulo logarithms].  A detailed
analysis shows that the asymptotic fall-off takes the form
$F_2(Q^2)/F_1(Q^2) \sim \log^2 Q^2/Q^2$~\cite{Belitsky:2002kj}.

A model~\cite{Brodsky:2003pw} incorporating the leading-twist
perturbative QCD prediction is consistent with the JLab polarization
transfer data~\cite{Jones:1999rz} for the ratio of proton Pauli and
Dirac form factors. Our analysis can also be extended to study the
spin structure of scattering amplitudes at large transverse momentum
and other processes which are dependent on the scaling and orbital
angular momentum structure of light-front wavefunctions. Recently,
Afanasev, Carlson, Chen, Vanderhaeghen, and I have shown that the
interfering two-photon exchange contribution to elastic
electron-proton scattering, including inelastic intermediate states,
can account for the discrepancy between Rosenbluth and polarization
data~\cite{Chen:2004tw}.

A crucial prediction of models for proton form factors is the
relative phase of the timelike form factors, since this can be
measured from the proton single spin symmetries in $e^+ e^- \to p
\bar p$ or $p \bar p \to \ell \bar \ell$~\cite{Brodsky:2003gs}. The
Zemach radius of the proton is known to better than 2\% from the
comparison of hydrogen and muonium hyperfine splittings; this
constraint needs to be incorporated into any
analysis~\cite{Brodsky:2004ck}.

\section{Applicability of PQCD and Conformal Symmetry to Hard Exclusive
Processes}
\label{sec:3}

The PQCD/conformal symmetry predictions for hadron form factors are
leading-twist predictions.  The only mass parameter is the QCD
scale, so the power-law predictions must be relevant -- up to
logarithms -- even in the few GeV domain.  Note also that the same
PQCD couplings which enter hard exclusive reactions are tested in
DGLAP evolution even at small $Q^2.$ As noted above, the dimensional
counting rules for form factors and exclusive processes have also
been derived for conformal QCD using the AdS/CFT
correspondence~\cite{Polchinski:2001tt,Brodsky:2003px}.

In fact, there have been a remarkable number of empirical successes
of PQCD predictions, including the scaling and angular dependence of
$\gamma \gamma \to \pi^+ \pi^-$, pion photoproduction, vector meson
electroproduction, and the photon-to-pion transition form factor.
A particularly dramatic example is deuteron photodisintegration
which satisfies the predicted scaling law $[s^{11} {d\sigma\over
dt}(\gamma d \to p n)\sim {\rm const}]$ at large $p_\perp$ and fixed
CM angle~\cite{Rossi:2004qm} to remarkable high precision.
Perturbative QCD predicts that only the small compact part of the
light-front wavefunctions enter exclusive hard scattering processes,
and that these hadronic fluctuations have diminished interactions in
a nuclear target~\cite{Brodsky:1988xz}. Evidence for QCD color
transparency has been observed for quasi-elastic
photoproduction~\cite{Dutta:2003mk} and proton-proton
scattering~\cite{Aclander:2004zm}. In general, the PQCD scaling
behavior can be modulated by resonances and heavy quark threshold
phenomena~\cite{Brodsky:1987xw} which can cause dramatic spin
correlations~\cite{Court:1986dh} as well as novel color transparency
effects~\cite{Brodsky:1988xz,Aclander:2004zm}. The approach to
scaling in pion photoproduction: $[s^{7} {d\sigma\over dt}(\gamma p
\to n \pi^+)\sim {\rm const}]$ and evidence for structure due to the
strangeness threshold has recently been studied at Jefferson
Laboratory~\cite{Zhu:2004dy}.

Leading-order perturbative QCD predicts the empirical scaling of
form factors and other hard exclusive amplitudes, but it typically
underestimate the normalization  The normalization of theoretical
prediction involves questions of the shape of the hadron
distribution amplitudes, the proper scale for the running
coupling~\cite{Brodsky:1997dh} as well as higher order corrections.
In fact, as noted above, in the AdS/CFT analysis, hard scattering
amplitudes which are normally of order $\alpha_s^p$ in PQCD appear
as order $\alpha_s^{p/2}$ in the nonperturbative theory
~\cite{Maldacena:1997re,Rey:1998ik}.

The observation of conformal scaling behavior~\cite{Brodsky:1976rz}
in exclusive deuteron processes such as deuteron
photoproduction~\cite{Rossi:2004qm}  and the deuteron form
factor~\cite{Arnold:1975dd} is particularly interesting. For
example, at high $Q^2$ the deuteron form factor is sensitive to
wavefunction configurations where all six quarks overlap within an
impact separation $b_{\perp i} < {\cal O} (1/Q).$ In general, the
six-quark wavefunction of a deuteron is a mixture of five different
color-singlet states. The dominant color configuration at large
distances corresponds to the usual proton-neutron bound state.
However at small impact space separation, all five Fock
color-singlet components eventually acquire equal weight, i.e., the
deuteron wavefunction evolves to 80\%\ ``hidden color.'' The
derivation of the evolution equation for the deuteron distribution
amplitude and its leading anomalous dimension $\gamma$ is given in
Ref.~\cite{Brodsky:1983vf}. The relatively large normalization of
the deuteron form factor observed at large
$Q^2$~\cite{Farrar:1991qi}, as well as the presence of two mass
scales in the scaling behavior of the reduced deuteron form
factor~\cite{Brodsky:1976rz} $f_d(Q^2)= F_d(Q^2)/F^2(Q^2/4)$
suggests sizable hidden-color contributions in the deuteron
wavefunction.

\vspace{.5in} \centerline{\bf Acknowledgements}

\noindent It is a pleasure to thank  Omar Benhar, Adelchi Fabrocini,
and Rocco Schiavilla, the organizers of the Electron-Nucleus
Scattering VIII meeting for their hospitality in Elba. I also thank
my collaborators, particularly Carl Carlson, Guy de Teramond, Rikard
Enberg, Paul Hoyer, Dae Sung Hwang, Gunnar Ingelman, Volodya
Karmanov, Joerg Raufeisen, and Ivan Schmidt. This work was supported
by the Department of Energy, contract No. DE-AC02-76SF00515.


\begin{thebibliography}{}


\bibitem{Bjorken:1968dy}
J.~D.~Bjorken,
Phys.\ Rev.\  {\bf 179}, 1547 (1969).



\bibitem{Bloom:1969kc}
E.~D.~Bloom {\it et al.},
Phys.\ Rev.\ Lett.\  {\bf 23}, 930 (1969).

\bibitem{Zeller:2001hh}
G.~P.~Zeller {\it et al.}  [NuTeV Collaboration],
%
Phys.\ Rev.\ Lett.\  {\bf 88}, 091802 (2002)
[Erratum-ibid.\  {\bf 90}, 239902 (2003)]
[arXiv:hep-ex/0110059].


\bibitem{Brodsky:2002ue}
S.~J.~Brodsky, P.~Hoyer, N.~Marchal, S.~Peigne and F.~Sannino,
Phys.\ Rev.\ D {\bf 65}, 114025 (2002)
[arXiv:hep-ph/0104291].


\bibitem{Brodsky:2002cx}
S.~J.~Brodsky, D.~S.~Hwang and I.~Schmidt,
Phys.\ Lett.\ B {\bf 530}, 99 (2002) [arXiv:hep-ph/0201296].

\bibitem{Brodsky:1989qz}
S.~J.~Brodsky and H.~J.~Lu,
Phys.\ Rev.\ Lett.\  {\bf 64}, 1342 (1990).



\bibitem{Brodsky:2004hi}
S.~J.~Brodsky, R.~Enberg, P.~Hoyer and G.~Ingelman,
arXiv:hep-ph/0409119.

\bibitem{Belitsky:2002sm}
A.~V.~Belitsky, X.~Ji and F.~Yuan,
Nucl.\ Phys.\ B {\bf 656}, 165 (2003) [arXiv:hep-ph/0208038].

\bibitem{Collins:2002kn}
J.~C.~Collins,
Phys.\ Lett.\ B {\bf 536}, 43 (2002)
[arXiv:hep-ph/0204004].


\bibitem{Brodsky:2004qa}
S.~J.~Brodsky, I.~Schmidt and J.~J.~Yang,
SLAC-PUB-9677.

\bibitem{cdkm}
J.~Carbonell, B.~Desplanques, V.~A.~Karmanov, and J.~F.~Mathiot,
Phys.\ Rep.\  {\bf 300}, 215 (1998) [arXiv:nucl-th/9804029].


\bibitem{Tomboulis:jn}
E.~Tomboulis,
Phys.\ Rev.\ D {\bf 8}, 2736 (1973).

\bibitem{Srivastava:2000cf}
P.~P.~Srivastava and S.~J.~Brodsky,
Phys.\ Rev.\ D {\bf 64}, 045006 (2001) [arXiv:hep-ph/0011372].

\bibitem{Brodsky:1998hn}
S.~J.~Brodsky and D.~S.~Hwang,
Nucl.\ Phys.\ B {\bf 543}, 239 (1999) [arXiv:hep-ph/9806358].

\bibitem{Brodsky:2000xy}
S.~J.~Brodsky, M.~Diehl and D.~S.~Hwang,
Nucl.\ Phys.\ B {\bf 596}, 99 (2001) [arXiv:hep-ph/0009254].

\bibitem{Diehl:2000xz}
M.~Diehl, T.~Feldmann, R.~Jakob and P.~Kroll,
Nucl.\ Phys.\ B {\bf 596}, 33 (2001) [Erratum-ibid.\ B {\bf 605}, 647 (2001)]
[arXiv:hep-ph/0009255].

\bibitem{Brodsky:1971zh}
S.~J.~Brodsky, F.~E.~Close and J.~F.~Gunion,
Phys.\ Rev.\ D {\bf 5}, 1384 (1972).


\bibitem{Brodsky:1973hm}
S.~J.~Brodsky, F.~E.~Close and J.~F.~Gunion,
Phys.\ Rev.\ D {\bf 8}, 3678 (1973).

\bibitem{Raufeisen:2004dg}
J.~Raufeisen and S.~J.~Brodsky,
arXiv:hep-th/0408108.



\bibitem{Jaffe:1989jz}
R.~L.~Jaffe and A.~Manohar,
Nucl.\ Phys.\ B {\bf 337}, 509 (1990).

\bibitem{Ji:2002qa}
X.~Ji,
arXiv:hep-lat/0211016.

\bibitem{BD80}
S.~J.~Brodsky and S.~D.~Drell,
Phys.\ Rev.\ D {\bf 22}, 2236 (1980).

\bibitem{Brodsky:2003gk}
S.~J.~Brodsky,
arXiv:hep-th/0304106.

\bibitem{Brodsky:1988xz}
S.~J.~Brodsky and A.~H.~Mueller,
Phys.\ Lett.\ B {\bf 206}, 685 (1988).

\bibitem{Bertsch:1981py}
G.~Bertsch, S.~J.~Brodsky, A.~S.~Goldhaber and J.~F.~Gunion,
Phys.\ Rev.\ Lett.\  {\bf 47}, 297 (1981).

\bibitem{Frankfurt:1999tq}
L.~Frankfurt, G.~A.~Miller and M.~Strikman,
Found.\ Phys.\  {\bf 30}, 533 (2000) [arXiv:hep-ph/9907214].

\bibitem{Nikolaev:2000sh}
N.~N.~Nikolaev, W.~Schafer and G.~Schwiete,
Phys.\ Rev.\ D {\bf 63}, 014020 (2001) [arXiv:hep-ph/0009038].

\bibitem{Aitala:2000hc}
E.~M.~Aitala {\it et al.}  [E791 Collaboration],
Phys.\ Rev.\ Lett.\  {\bf 86}, 4773 (2001) [arXiv:hep-ex/0010044].

\bibitem{Aitala:2000hb}
E.~M.~Aitala {\it et al.}  [E791 Collaboration],
Phys.\ Rev.\ Lett.\  {\bf 86}, 4768 (2001) [arXiv:hep-ex/0010043].

\bibitem{Gronberg:1998fj}
J.~Gronberg {\it et al.}  [CLEO Collaboration],
Phys.\ Rev.\ D {\bf 57}, 33 (1998) [arXiv:hep-ex/9707031].

\cite{Lepage:1979zb}
\bibitem{Lepage:1979zb}
G.~P.~Lepage and S.~J.~Brodsky,
Phys.\ Lett.\ B {\bf 87}, 359 (1979).

\bibitem{Efremov:1978rn}
A.~V.~Efremov and A.~V.~Radyushkin,
Theor.\ Math.\ Phys.\  {\bf 42}, 97 (1980) [Teor.\ Mat.\ Fiz.\ {\bf 42}, 147 (1980)].

\bibitem{Lepage:1980fj}
G.~P.~Lepage and S.~J.~Brodsky,
{\em Phys.\ Rev.\ D} {\bf 22}, 2157 (1980).


\bibitem{Brodsky:1980pb}
S.~J.~Brodsky, P.~Hoyer, C.~Peterson and N.~Sakai,
Phys.\ Lett.\ B {\bf 93}, 451 (1980).

\bibitem{Franz:2000ee}
M.~Franz,~V.~Polyakov and K.~Goeke,
Phys.\ Rev.\ D {\bf 62}, 074024 (2000)
[arXiv:hep-ph/0002240].



\bibitem{Brodsky:1984nx}
S.~J.~Brodsky, J.~C.~Collins, S.~D.~Ellis, J.~F.~Gunion and A.~H.~Mueller,
DOE/ER/40048-21 P4
{\it Proc. of 1984 Summer Study on the SSC, Snowmass, CO, Jun 23 - Jul
13, 1984}


\bibitem{Harris:1995jx}
B.~W.~Harris, J.~Smith and R.~Vogt,
Nucl.\ Phys.\ B {\bf 461}, 181 (1996)
[arXiv:hep-ph/9508403].


\bibitem{Anjos:2001jr}
J.~C.~Anjos, J.~Magnin and G.~Herrera,
Phys.\ Lett.\ B {\bf 523}, 29 (2001)
[arXiv:hep-ph/0109185].


\bibitem{Ocherashvili:2004hi}
A.~Ocherashvili {\it et al.}  [SELEX Collaboration],
arXiv:hep-ex/0406033.


\bibitem{Vogt:1995tf}
R.~Vogt and S.~J.~Brodsky,
Phys.\ Lett.\ B {\bf 349}, 569 (1995)
[arXiv:hep-ph/9503206].


\bibitem{Brodsky:1997fj}
S.~J.~Brodsky and M.~Karliner,
Phys.\ Rev.\ Lett.\  {\bf 78}, 4682 (1997)
[arXiv:hep-ph/9704379].


\bibitem{Brodsky:2001yt}
S.~J.~Brodsky and S.~Gardner,
Phys.\ Rev.\ D {\bf 65}, 054016 (2002)
[arXiv:hep-ph/0108121].


\bibitem{Parisi:zy}
G.~Parisi,
{\em Phys.\ Lett.\ B} {\bf 39}, 643 (1972).

\bibitem{Brodsky:2000cr}
S.~J.~Brodsky, E.~Gardi, G.~Grunberg and J.~Rathsman,
Phys.\ Rev.\ D {\bf 63}, 094017 (2001)
[arXiv:hep-ph/0002065].

\bibitem{Rathsman:2001xe}
J.~Rathsman,
in {\it Proc. of the 5th International Symposium on Radiative Corrections
(RADCOR 2000) }
ed. Howard E. Haber, arXiv:hep-ph/0101248.


\bibitem{Grunberg:2001bz}
G.~Grunberg,
JHEP {\bf 0108}, 019 (2001)
[arXiv:hep-ph/0104098].


\bibitem{Banks:1981nn}
T.~Banks and A.~Zaks,
Nucl.\ Phys.\ B {\bf 196}, 189 (1982).


\bibitem{Brodsky:1985ve}
S.~J.~Brodsky, Y.~Frishman and G.~P.~Lepage,
{\em Phys.\ Lett.\ B} {\bf 167}, 347 (1986).

\bibitem{Brodsky:1984xk}
S.~J.~Brodsky, P.~Damgaard, Y.~Frishman and G.~P.~Lepage,
{\em Phys.\ Rev.\ D} {\bf 33}, 1881 (1986).


\bibitem{Braun:2003rp}
V.~M.~Braun, G.~P.~Korchemsky and D.~Muller,
Prog.\ Part.\ Nucl.\ Phys.\  {\bf 51}, 311 (2003)
[arXiv:hep-ph/0306057].



\bibitem{Braun:1999te}
V.~M.~Braun, S.~E.~Derkachov, G.~P.~Korchemsky and A.~N.~Manashov,
{\em Nucl.\ Phys.\ B} {\bf 553}, 355 (1999)
[arXiv:hep-ph/9902375].




\bibitem{Brodsky:1994eh}
S.~J.~Brodsky and H.~J.~Lu,
{\em Phys.\ Rev.\ D} {\bf 51}, 3652 (1995) [arXiv:hep-ph/9405218].

\bibitem{Maldacena:1997re}
J.~M.~Maldacena,
{\em Adv.\ Theor.\ Math.\ Phys.}\  {\bf 2}, 231 (1998) [{\em Int.\
J.\ Theor.\ Phys.}\ {\bf 38}, 1113 (1999)] [arXiv:hep-th/9711200].

\bibitem{Polchinski:2001tt}
J.~Polchinski and M.~J.~Strassler,
{\em Phys.\ Rev.\ Lett.}\  {\bf 88}, 031601 (2002)
[arXiv:hep-th/0109174].

\bibitem{Brower:2002er}
R.~C.~Brower and C.~I.~Tan,
{\em Nucl.\ Phys.\ B} {\bf 662}, 393 (2003)
[arXiv:hep-th/0207144].

\bibitem{Andreev:2002aw}
O.~Andreev,
{\em Phys.\ Rev.\ D} {\bf 67}, 046001 (2003)
[arXiv:hep-th/0209256].

\bibitem{Brodsky:1973kr}
S.~J.~Brodsky and G.~R.~Farrar,
{\em Phys.\ Rev.\ Lett.}\  {\bf 31}, 1153 (1973).

\bibitem{Matveev:ra}
V.~A.~Matveev, R.~M.~Muradian and A.~N.~Tavkhelidze,
{\em Lett.\ Nuovo Cim.}\  {\bf 7}, 719 (1973).

\bibitem{Brodsky:1974vy}
S.~J.~Brodsky and G.~R.~Farrar,
{\em Phys.\ Rev.\ D} {\bf 11}, 1309 (1975).

\bibitem{Brodsky:2002st}
S.~J.~Brodsky,
Published in {\it Newport News 2002, Exclusive processes at high momentum
transfer 1-33.}
[arXiv:hep-ph/0208158.]


\bibitem{Brodsky:1994kg}
S.~J.~Brodsky, M.~Burkardt and I.~Schmidt,
{\em Nucl.\ Phys.\ B} {\bf 441}, 197 (1995)
[arXiv:hep-ph/9401328].


\bibitem{Bloom:1971ye}
E.~D.~Bloom and F.~J.~Gilman,
Phys.\ Rev.\ D {\bf 4}, 2901 (1971).


\bibitem{Rey:1998ik}
S.~J.~Rey and J.~T.~Yee,
{\em Eur.\ Phys.\ J.\ C} {\bf 22}, 379 (2001)
[arXiv:hep-th/9803001].

\bibitem{'tHooft:1973jz}
G.~'t Hooft,
{\em Nucl.\ Phys.\ B} {\bf 72}, 461 (1974).



\bibitem{Gunion:1973ex}
J.~F.~Gunion, S.~J.~Brodsky and R.~Blankenbecler,
{\em Phys.\ Rev.\ D} {\bf 8}, 287 (1973).

\bibitem{Sivers:1975dg}
D.~W.~Sivers, S.~J.~Brodsky and R.~Blankenbecler,
{\em Phys.\ Rept.}\  {\bf 23}, 1 (1976).

\bibitem{Blankenbecler:1973kt}
R.~Blankenbecler {\em et al.},
{\em Phys.\ Rev.\ D} {\bf 8}, 4117 (1973).



\bibitem{Brodsky:2003px}
S.~J.~Brodsky and G.~F.~de Teramond,
Phys.\ Lett.\ B {\bf 582}, 211 (2004)
[arXiv:hep-th/0310227].


\bibitem{deTeramond:2004qd}
G.~F.~de Teramond and S.~J.~Brodsky,
arXiv:hep-th/0409074.

\bibitem{Ji:bw}
X.~D.~Ji, F.~Yuan and J.~P.~Ma,
{\em Phys.\ Rev.\ Lett.}\  {\bf 90}, 241601 (2003).


\bibitem{Brodsky:2003pw}
S.~J.~Brodsky, J.~R.~Hiller, D.~S.~Hwang and V.~A.~Karmanov,
Phys.\ Rev.\ D {\bf 69}, 076001 (2004)
[arXiv:hep-ph/0311218].


\cite{Ji:2003fw}
\bibitem{Ji:2003fw}
X.~d.~Ji, J.~P.~Ma and F.~Yuan,
Phys.\ Rev.\ Lett.\  {\bf 90}, 241601 (2003) [arXiv:hep-ph/0301141].

\bibitem{Belitsky:2002kj}
A.~V.~Belitsky, X.~D.~Ji and F.~Yuan,
Phys.\ Rev.\ Lett.\  {\bf 91}, 092003 (2003)
[arXiv:hep-ph/0212351].


\bibitem{Jones:1999rz}
M.~K.~Jones {\it et al.}  [Jefferson Lab Hall A Collaboration],
Phys.\ Rev.\ Lett.\  {\bf 84}, 1398 (2000)
[arXiv:nucl-ex/9910005].



\bibitem{Chen:2004tw}
Y.~C.~Chen, A.~Afanasev, S.~J.~Brodsky, C.~E.~Carlson and M.~Vanderhaeghen,
arXiv:hep-ph/0403058.


\bibitem{Brodsky:2003gs}
S.~J.~Brodsky, C.~E.~Carlson, J.~R.~Hiller and D.~S.~Hwang,
Phys.\ Rev.\ D {\bf 69}, 054022 (2004)
[arXiv:hep-ph/0310277].


\bibitem{Brodsky:2004ck}
S.~J.~Brodsky, C.~E.~Carlson, J.~R.~Hiller and D.~S.~Hwang,
arXiv:hep-ph/0408131.

\bibitem{Rossi:2004qm}
P.~Rossi {\it et al.}  [CLAS Collaboration],
arXiv:hep-ph/0405207.

\bibitem{Dutta:2003mk}
D.~Dutta {\it et al.}  [Jefferson Lab E940104 Collaboration],
Phys.\ Rev.\ C {\bf 68}, 021001 (2003) [arXiv:nucl-ex/0305005].


\bibitem{Aclander:2004zm}
J.~Aclander {\it et al.},
Phys.\ Rev.\ C {\bf 70}, 015208 (2004) [arXiv:nucl-ex/0405025].

\bibitem{Brodsky:1987xw}
S.~J.~Brodsky and G.~F.~de Teramond,
Phys.\ Rev.\ Lett.\  {\bf 60}, 1924 (1988).

\bibitem{Court:1986dh}
R.~Court {\it et al.},
Phys.\ Rev.\ Lett.\  {\bf 57}, 507 (1986).

\bibitem{Zhu:2004dy}
L.~Y.~Zhu  [Jefferson Lab E94-104 Collaboration],
arXiv:nucl-ex/0409018.


\bibitem{Brodsky:1997dh}
S.~J.~Brodsky, C.~R.~Ji, A.~Pang and D.~G.~Robertson,
Phys.\ Rev.\ D {\bf 57}, 245 (1998)
[arXiv:hep-ph/9705221].

\bibitem{Brodsky:1976rz}
S.~J.~Brodsky and B.~T.~Chertok,
The Continuity Of
%
Phys.\ Rev.\ D {\bf 14}, 3003 (1976).

\bibitem{Arnold:1975dd}
R.~G.~Arnold {\it et al.},
Phys.\ Rev.\ Lett.\  {\bf 35}, 776 (1975).

\bibitem{Brodsky:1983vf}
S.~J.~Brodsky, C.~R.~Ji and G.~P.~Lepage,
Phys.\ Rev.\ Lett.\  {\bf 51}, 83 (1983).

\bibitem{Farrar:1991qi}
G.~R.~Farrar, K.~Huleihel and H.~y.~Zhang,
Phys.\ Rev.\ Lett.\  {\bf 74}, 650 (1995).


\end{thebibliography}
\end{document}